\documentclass[11pt,a4paper,final]{iopart}
\usepackage[utf8]{inputenc}
\usepackage{iopams}
\usepackage{graphicx}
\usepackage[breaklinks=true,colorlinks=true,linkcolor=blue,urlcolor=blue,citecolor=blue]{hyperref}
\usepackage[labelfont=bf]{caption}
\usepackage[usenames,dvipsnames,svgnames,table]{xcolor}
\usepackage{enumerate}
\usepackage{changepage} 
\usepackage{multirow}	
\usepackage{soul}       
\usepackage{ulem}       
\usepackage{xcolor}     
\usepackage[hang]{footmisc}	
\usepackage{appendix}
\usepackage{comment}		

\setstcolor{red}    

\begin{document}
\title{\textcolor{black}{Probing the effect on student conceptual understanding due to  
a forced mid-semester transition to online teaching}}

\author{Emanuela Carleschi, Anna Chrysostomou, Alan S. Cornell} 
\address{Department of Physics, University of Johannesburg, PO Box 524, Auckland Park 2006, South 
Africa}
\ead{ecarleschi@uj.ac.za, annachrys97@gmail.com, acornell@uj.ac.za} 

\author{Wade Naylor}
\address{Immanuel Lutheran College, PO Box 5025, Maroochydore 4558, Australia \&}
\address{Department of Physics, University of Johannesburg, PO Box 524, Auckland Park 2006, South 
Africa}
\ead{naylorw@immanuel.qld.edu.au}

\begin{abstract}
The Force Concept Inventory (FCI) can be used as an assessment tool to measure conceptual gains in a cohort of students. \textcolor{black}{The FCI uses a conceptions/``misconceptions" lens rather than a context dependent perspective, such as ``knowledge-in-pieces".} In this study it was given to first year students ($N=256$ students) pre- and post-mechanics lectures, at the University of Johannesburg. From these results we examine the effect of switching mid-semester from traditional classes to online classes, as imposed by the COVID-19 lockdown in South Africa. Overall results indicate no appreciable difference of gain, when bench-marked against previous studies using this assessment tool. When compared with $2019$ grades, the $2020$ semester grades do not appear to be greatly affected. Furthermore, statistical analyses also indicate a gender difference in mean gains in favour of females at the $95\%$ significance level (for paired data, $N=48$).
\end{abstract}

\vspace{2pc}
\noindent{\it Keywords}: Physics Education; Large Cohort Courses; Online Teaching; Force Concept Inventory; FCI.

\maketitle
\tableofcontents


\section{Introduction}\label{sec:1}

\par Many studies through the years have advocated for various changes to teaching 
pedagogy away from the so-called traditional lecturing pedagogy, such as flipped 
classrooms, and peer-assessment \cite{Heller:1992, Fagen:2002, Cah:2004, Smith:2009, 
Watkins:2013, Freeman:2014}, to name but a few. Many of these studies have used 
assessment tools, such as the Force Concept Inventory (FCI) \cite{Hall:1985,Hest:1992, 
Hest:1998}, to assess the efficacy of these changes, where such drastic changes have 
been regarded over many years of studies. In the following subsections we discuss some of the 
history behind the FCI, the South African context and the plan for this article.

\subsection{The Force Concept Inventory: Some History}\label{sec:1a}
\par Any physics department that decides to use an assessment tool has to go 
through the decision making process of choosing which kind of assessment of the many possible and
the reasons why \cite{Madsen:2017}. As discussed in Ref. \cite{Madsen:2017}, one must choose a 
particular topic to focus on, so we decided to look at conceptual issues in classical mechanics, 
because this is one of those core foundations that is taught across not just physics but also 
engineering and teaching degrees at the University of Johannesburg (UJ). We have chosen to use 
the FCI as the type of assessment for the understanding of mechanics concepts by first year 
students. The current version of the FCI was released in 1995 and is known as v95. This version 
has 30 questions, fewer ambiguities, and a smaller likelihood of false positives than the 
original version \cite{Hake:1998}. The original version had 29 questions \cite{Hest:1992}, which 
was a revision of an earlier test called the Mechanics Diagnostic Test (MDT) \cite{Hall:1985}.

\par One motivation for choosing the FCI is that it has not only been used by 
many other institutions internationally \cite{Hake:1998}, but has also been under immense 
scrutiny over the decades preceding it. For example, the issue of false positives has further 
been investigated by Yasuda et al. \cite{Yasuda:2018}, where systematic errors generated by 
false positives were found to be statistically significant for questions: Q.6, Q.7, Q.16 and not 
Q.5 ($N=1110$). However, the authors' recommendation was not to change the question structure 
given the fact that the current version (v95) has been extensively used in the literature. Other 
issues with how the FCI questions - and other assessments - are written in terms of ``none of the
above'' (NOTA) and ``zero'' distractors have also been discussed in Ref. \cite{Devore:2016} (and 
references therein). There have also been extensive studies on gender, as we discuss in Sec. 
\ref{sec:3gend}, and finally the FCI has been translated to a number of languages. We come back 
to this issue for South Africa in the next subsection.

\par Given the above considerations we may consider the FCI as a \textit{de facto} standard for 
assessing the conceptual basis of Newtonian mechanics, for which the gain between pre- and 
post-test results serves as a benchmark for standard performance on a course. The v95 version of 
the FCI was given at the start of a course on mechanics as a ``pre-test" and then at the end in 
the form of a ``post-test". The normalised gain $G$ \cite{Hake:1998} is defined as:
\begin{equation}
G = {\langle \% S_f\rangle  - \langle \% S_i \rangle \over 100 - \langle \% S_i \rangle 
},
\label{G}
\end{equation}
\noindent
where $\% S_f$ and $\% S_i$ are the final and initial scores respectively. The reported gain 
associated with students enrolled in an introductory mechanics course is approximately $G = 25 
\%$ \cite{Hake:1998,Coletta:2005}. With this established tool and quantifiable indicator of 
performance, we may probe the question of how a forced transition from in-person to online 
teaching affected the conceptual understanding of Newtonian mechanics in a first-year cohort at 
what might be considered a relatively low-resource institution for higher-education.

To further enrich our analysis of students' comprehension, we compare student performance in the 
pre- and post-tests on a question-by-question basis. Such in-depth analyses have recently been 
performed in Refs. \cite{Alinea:2017,Alinea:2015,Alinea:2020}, where a breakdown of the type of 
response to individual questions can lead to a polarisation of a correct answer and one 
predominantly incorrect answer.\footnote{See Refs.~\cite{Martin_Blas:2010,Bani_Salameh:2016a, 
Bani_Salameh:2016b,Yasuda:2018} for a discussion of other ways to analyse and interpret 
individual question responses in the FCI.} This can also be useful in circumstances where only a 
pre- or post-test may be administered.

\subsection{South African Context}\label{sec:1b}

\par Less than one tenth of the students enrolled at the UJ are first language 
English speakers. Administering the FCI in English to these first year students might therefore 
introduce biases in the performance due to the proficiency in the English language rather than 
the actual understanding of the underlying physics concepts. This is, however, the only feasible 
way to conduct the FCI, because English is the most common spoken language in the country. The 
influence of English proficiency on the FCI performance within the specific South African context is 
something the authors shall investigate in future studies.

\par As for the relevance of the FCI within the South African context specifically, it is worthwhile taking a closer look at the diagnostic 
reports on the learners' performance in the National Senior Certificate (i.e. their final year of high school) exams in the Physical Sciences 
paper \cite{DBEwebsite}, compiled by the National Department of Basic Education of South Africa for the academic year 2019 
\cite{DBEreport2019} (which is the year of completion for the first year cohort of physics students under investigation in this study), as 
well as for 2020 and previous years \cite{DBEwebsite}. With regards to mechanics-specific questions, all the reports highlight recurrent 
challenges experienced by learners in properly identifying the direction of forces applied to objects, as well as drawing and labelling 
free-body diagrams correctly. This result is very interesting in light  of the findings we report in this work: that the questions in the FCI 
that specifically deal with forces - i.e. Q.5, Q.11, Q.13, Q.18, Q.29 and Q.30 - are those showing polarisation of the answers. This hints to 
the fact that, despite its intrinsic challenges, the FCI can be used as a good indicator of students' ``misconceptions" in the specific South 
African context.\footnote{\textcolor{black}{In this article we shall use the word ``misconceptions" to mean ``prior-conceptions" as we shall briefly elaborate on in Sec. \ref{sec:2}.}}

\par Whilst this study has been conducted for only one year's worth of data (where, as mentioned 
above, usual analyses of such changes in pedagogy are conducted over several years), we believe 
that the unprecedented nature of the 2020 lockdown experienced in South Africa in the wake of the
COVID-19 pandemic warranted study. This work may be considered then as a useful indicator of the 
short-term effects of a forced transition to online learning on first-year mechanics students. As
such, our study was conducted with a very diverse group of first year students enrolled at the 
Faculty of Engineering and the Built Environment (FEBE), at UJ, whose demographics 
(average 2015 - 2019) range as follows: African 92.8$\%$; White 3.8$\%$; Indian 2.3$\%$; Coloured
1.1$\%$  \cite{FEBE2019AnnualReport}. Yet for such a diverse background, the gains appear to 
have remained comparable to the bench-marked studies of the last three decades. 

\par In seeking to unpack the fact that an change in pedagogy did not affect the 
report gain ($G$) we shall look at a number of factors, including the previously 
studied Gender Gap \cite{Doktor:2008, Glasser:2008, Madsen:2013, Bates:2013, 
Coleatta:2013, Shapiro:2012, Alinea:2017}, where there has been a resulting gender 
difference in favour of males in the student performance on standardised 
assessments, such as the FCI in previous studies. However, our results indicate this
does not seem to be the case here. Another aspect at play here could be the 
increased peer scaffolding (along the lines of Mazur's peer evaluation 
\cite{Mazur:1997}) as there had been an increased reliance on discussion groups with
peers, due to the COVID lockdown. This persisted into the second semester, where a 
range of discussion boards, interactive tutorials, and WhatsApp groups were used. As
presented in Table \ref{table-marks-2019-2020-comp}, it can be seen that the scores 
for the average Semester 2 mark for both 2019 and 2020, which includes a combination
of coursework, practicals, and exams, were not appreciably different. The only 
noticeable difference was in the number of students who passed after the November 
exam, or the {\it course throughput}, where explained in the table caption, in 2020 
a slightly lower semester mark was used to gain entrance to the final exam.

\subsection{Plan}\label{sec:1c}
\par Given these motivations our paper shall be presented as follows: In section 
\ref{sec:2} we shall detail the methodology of our study, followed by the analysis 
tools and techniques used in this study in section \ref{sec:3}, along with 
supporting appendices, and finally we shall conclude in section \ref{sec:4}.

\begin{table}[ht]
\begin{center}
\begin{tabular}{|c|cc|}
\hline 
  & 2019 & 2020 \\ 
\hline  
\hline
no. of enrolled students (excluding cancellations) & 349 & 404\\
average semester mark ($\%$), all students & 54 & 63\\
average semester mark ($\%$), only students who qualified for exam & 58 & 66\\
$\%$ students qualified for written exam & 75 & 93\\
average exam mark ($\%$) & 48 & 50 \\
average course mark ($\%$), all students & 45 & 55 \\
average course mark ($\%$), only students who qualified for exam & 53 & 58 \\
course throughput ($\%$) & 49 & 67.5 \\
\hline  
\end{tabular}
\caption{Comparison of the 2019 and 2020 marks for engineering physics 1 in the second semester. Note that: 1) the course throughput is calculated after the main exam only, excluding the results of the supplementary exams; 2) In 2020 the entrance requirement for the exam was lowered to $30\%$ for the theory part of the course (instead of the usual $40\%$ in 2019 and previous years) in light of the COVID-19 pandemic.}
\label{table-marks-2019-2020-comp}
\end{center}
\end{table}

\section{Methodology}\label{sec:2}

\par {\color{black}{Our study sought to explore the ``misconceptions" in Newtonian mechanics carried by early undergraduates in South African institutions of higher education. To study this we have used a conceptions/``misconceptions" lens \cite{smith1994misconceptions} rather than a context dependent perspective such as “knowledge-in-pieces” (KiP) \cite{diSessa2018_KIP}.  This is done as the FCI is premised on the notion of ``(mis)conceptions" and that it is easier to use this terminology.}} The subjects for our testing were the 2020 first year cohort of engineering students at the University of Johannesburg. The class consisted of approximately $400$ students (only four students dropped out during the semester transition, from $404$ to $400$). The course was initially taught as a traditional lecture-based course, with a weekly online assessment, fortnightly tutorials, and fortnightly practicals (these being done in person in groups of approximately 30 students with graduate students acting as tutors and demonstrators of the practicals). The academic year had begun in early February of 2020, where the pre-mechanics course FCI test was conducted in late February on $N = 256$ students.\footnote{From a cohort of $400$ students $256$ students sat the FCI, where for pre- and post-test cases there were $144$ and $166$ students, respectively. After removing several redundant attempts, there were $256$ data subjects in total. Those who took both tests (paired) were $48$ in total.} We found that for this cohort (for paired data) $N=48$  the average gain was at $24\%$. We will
comment further on these results in Sec. \ref{sec:3means}.

\par First we should note that South Africa was placed in a hard lockdown in mid-March 2020, and 
teaching was switched within the period of a few weeks to a purely online format. Lectures were 
replaced with recorded video content, such that the number of hours of contact time were unchanged. To 
provide additional support, online platforms for engagement with students were employed (such as 
consultations using BlackBoard Collaborate Ultra, and WhatsApp discussion groups). As the easing of the
lockdown occurred towards the end of the mechanics course, a post-FCI test was possible to administer 
to a smaller voluntary group of students ($N=166$), of which $N=48$ had done both the pre-test and post
test.

\par The methodology employed to unpack this collected data relied primarily on standard statistical 
parameters, including the mean, standard deviation, percent differences, \textit{p}-values for the 
\textit{t}-test difference of means, and correlations through R \cite{Crawley:2014, Rprogram} and a 
spreadsheet.\footnote{Interested readers who wish to familiarise themselves with the basics of 
statistical analysis, including the \textit{t}-test, correlation, ANOVA, and measures of variation 
(e.g. standard deviation and standard error of the mean), among others, may want to consult Refs. 
\cite{Crawley:2014, Urdan:2010}.}

\par Using our data from this 2020 cohort in Sec. \ref{sec:3} we shall investigate: \begin{enumerate}
    \item student performance from pre-test to post-test, including an analysis of their performance in the pre- and post-tests via a question breakdown, see Sec. \ref{sec:3means},
    
    \item the existence of a polarisation effect in 6 particular questions \cite{Alinea:2015}, see Sec. \ref{sec:3pol},
    
    \item {\color{blue}{and}} a possible gender difference in the FCI for paired data, see Sec. \ref{sec:3gend}.
    
\end{enumerate}

\begin{figure}[ht]
\centering
\begin{minipage}{.5\textwidth}
  \centering
  \includegraphics[width=1.0\linewidth]{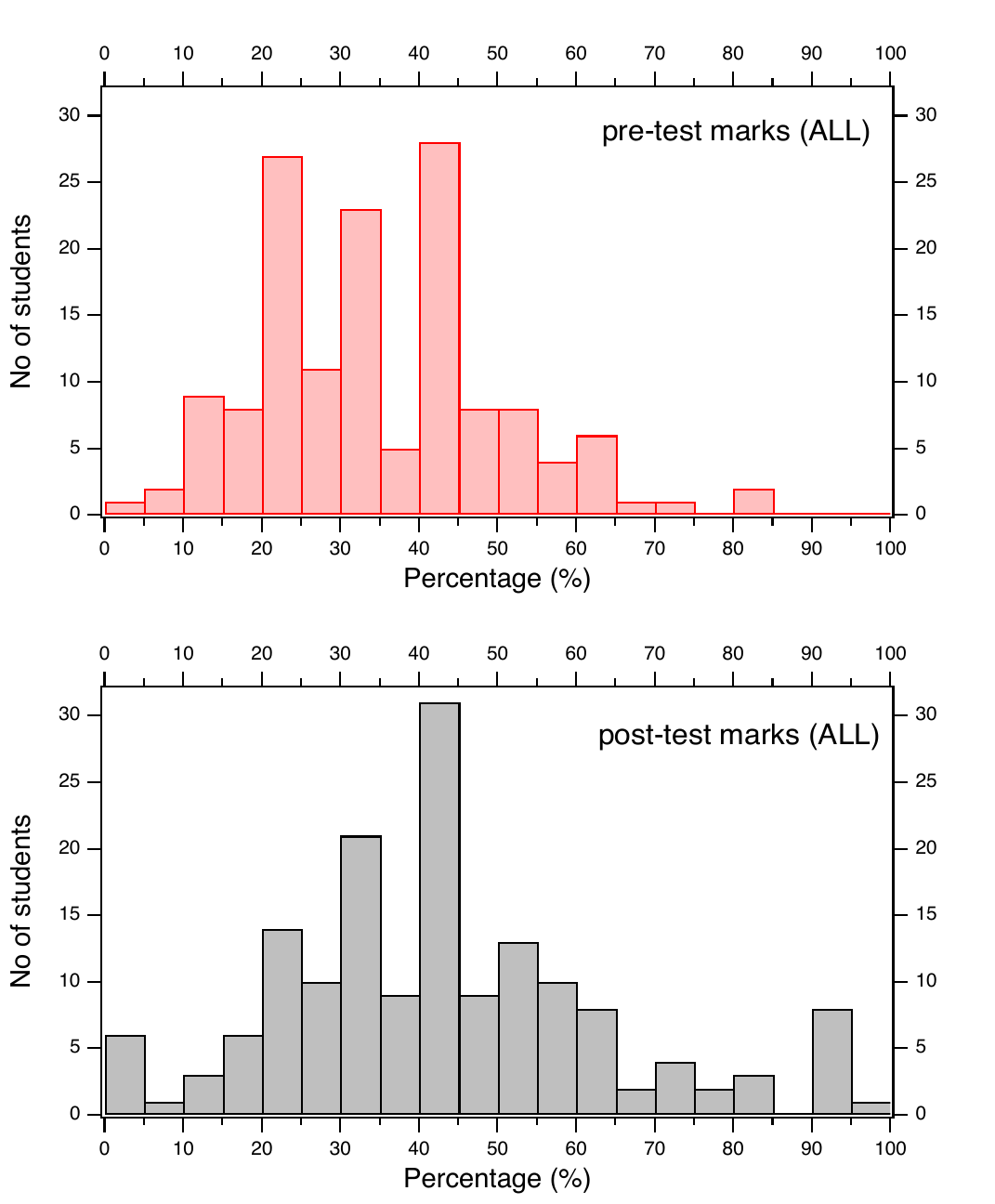}
\end{minipage}%
\begin{minipage}{.5\textwidth}
  \centering
  \includegraphics[width=1.0\linewidth]{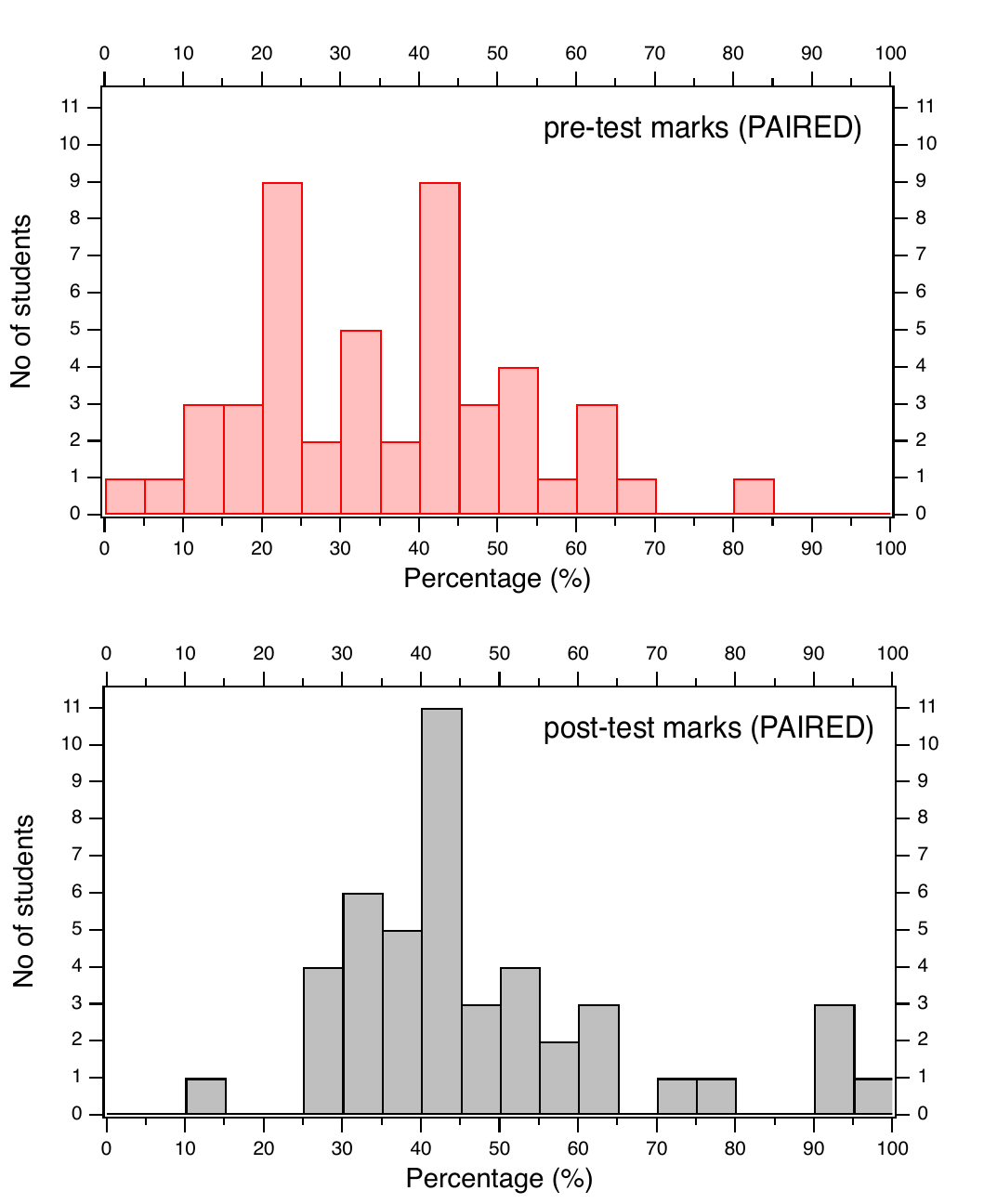}
\end{minipage}
\caption{A frequency histogram for the pre- and post-test data in total ($N=256$) and for the paired data ($N=48$).}
\label{fig:HistMarks}
\end{figure}
\begin{table}[ht]
\begin{center}
\begin{tabular}{|c|cccc|}
\hline 
  & Mean & SD & Min. & Max. \\ 
\hline  
\hline
pre-test ALL & \textcolor{black}{34.3} & 15.2 & 3 & 80\\
pre-test PAIRED & \textcolor{black}{34.7} & 17.2 & 3 & 80\\
post-test ALL & \textcolor{black}{44.1} & \textcolor{black}{22.8} & 0 & 100\\
post-test PAIRED & \textcolor{black}{50.8} & \textcolor{black}{22.4} & 10 & 100\\
\hline  
\end{tabular}
\caption{Mean, standard deviation (SD), minimum and maximum $\%$ marks as displayed in Fig. \ref{fig:HistMarks}. The paired data consisted of a subset ($N=48)$ of the $N=256$ who sat either the pre- or post test.}
\label{table-marks-stats}
\end{center}
\end{table}
\begin{figure}[ht]
    \centering
    \scalebox{0.55}{
    \includegraphics{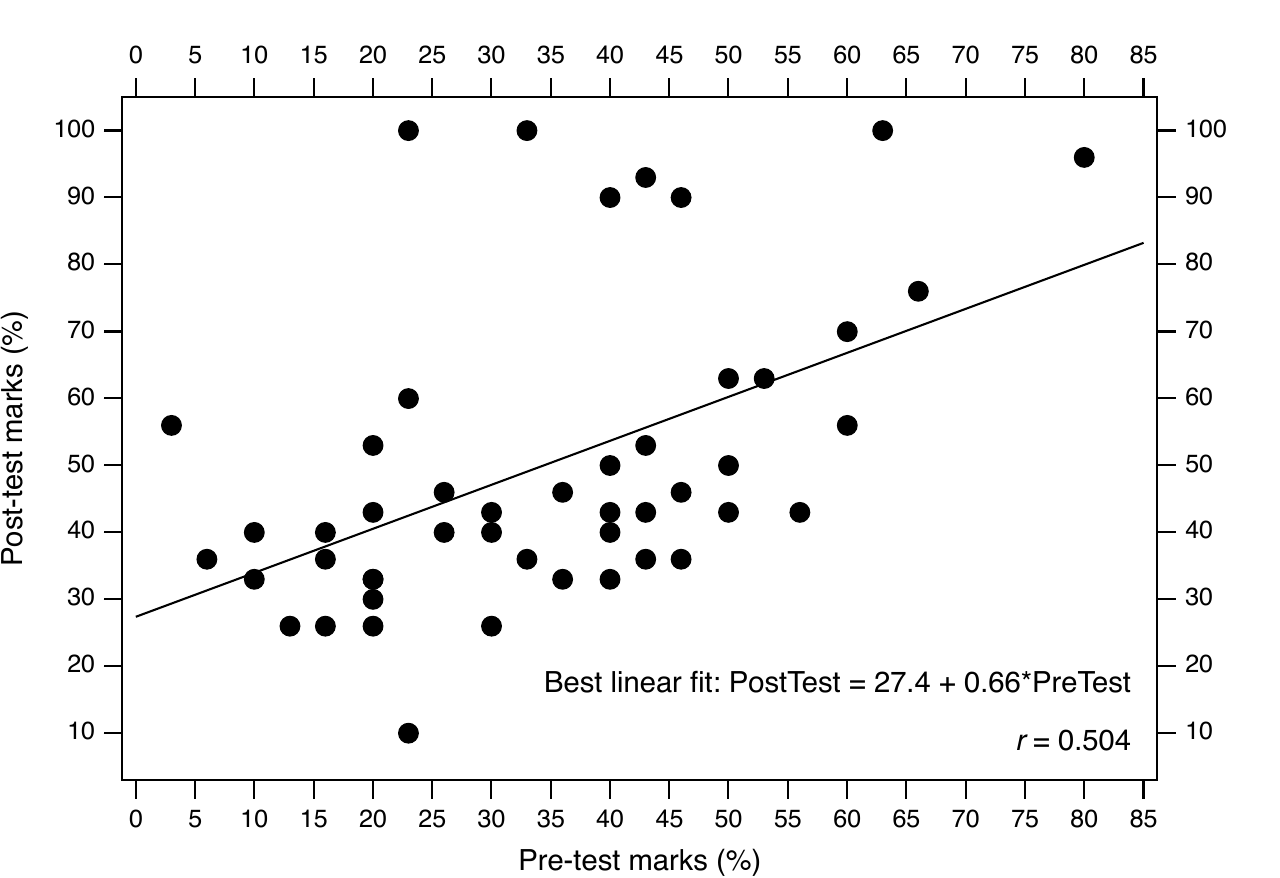}}
    \caption{Moderate positive correlation between pre- and post-tests, for paired data $N=48$.}
    \label{fig:correlation-marks}
\end{figure}
\begin{figure}[ht]
    \centering
    \scalebox{0.385}{
    \includegraphics{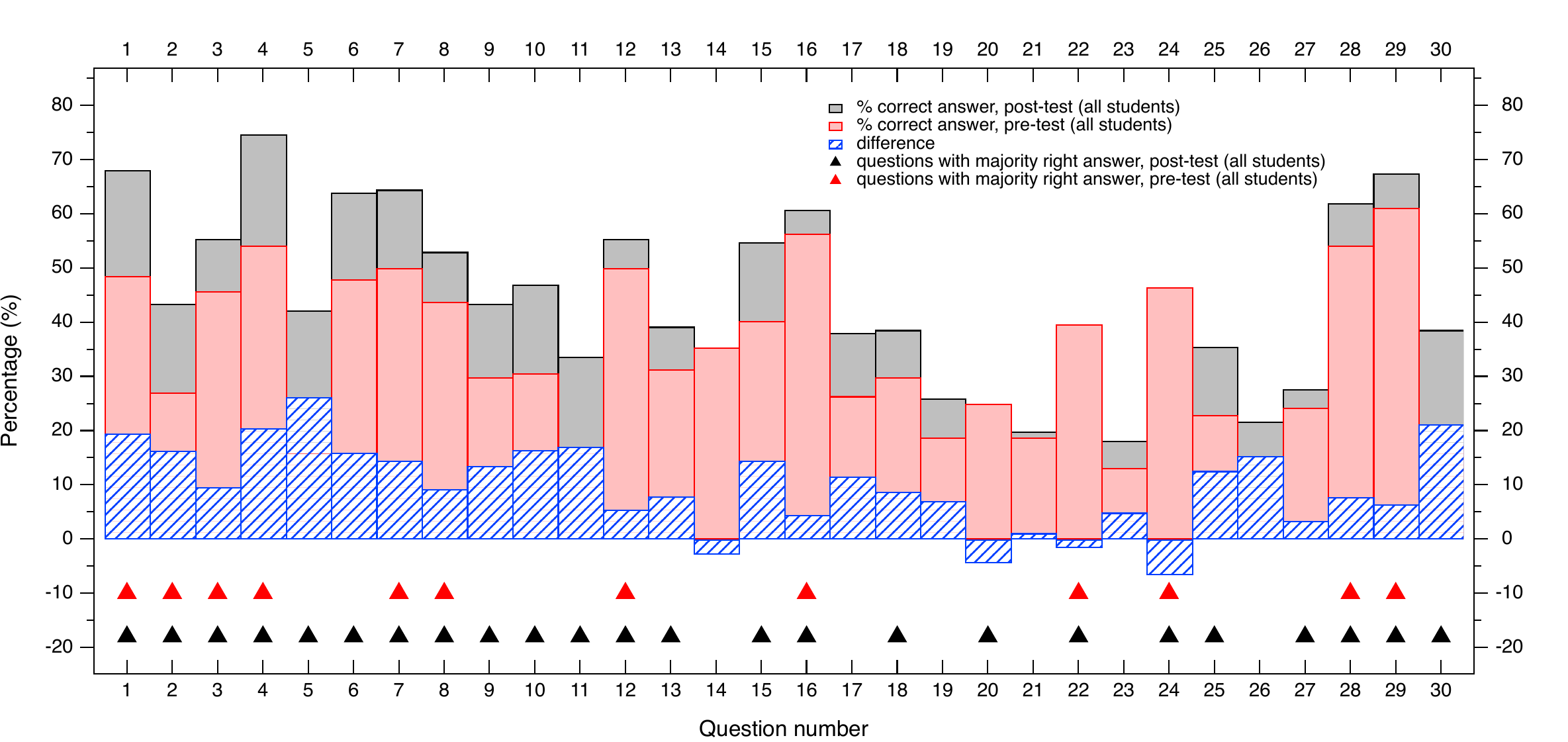}}
    \scalebox{0.385}{
    \includegraphics{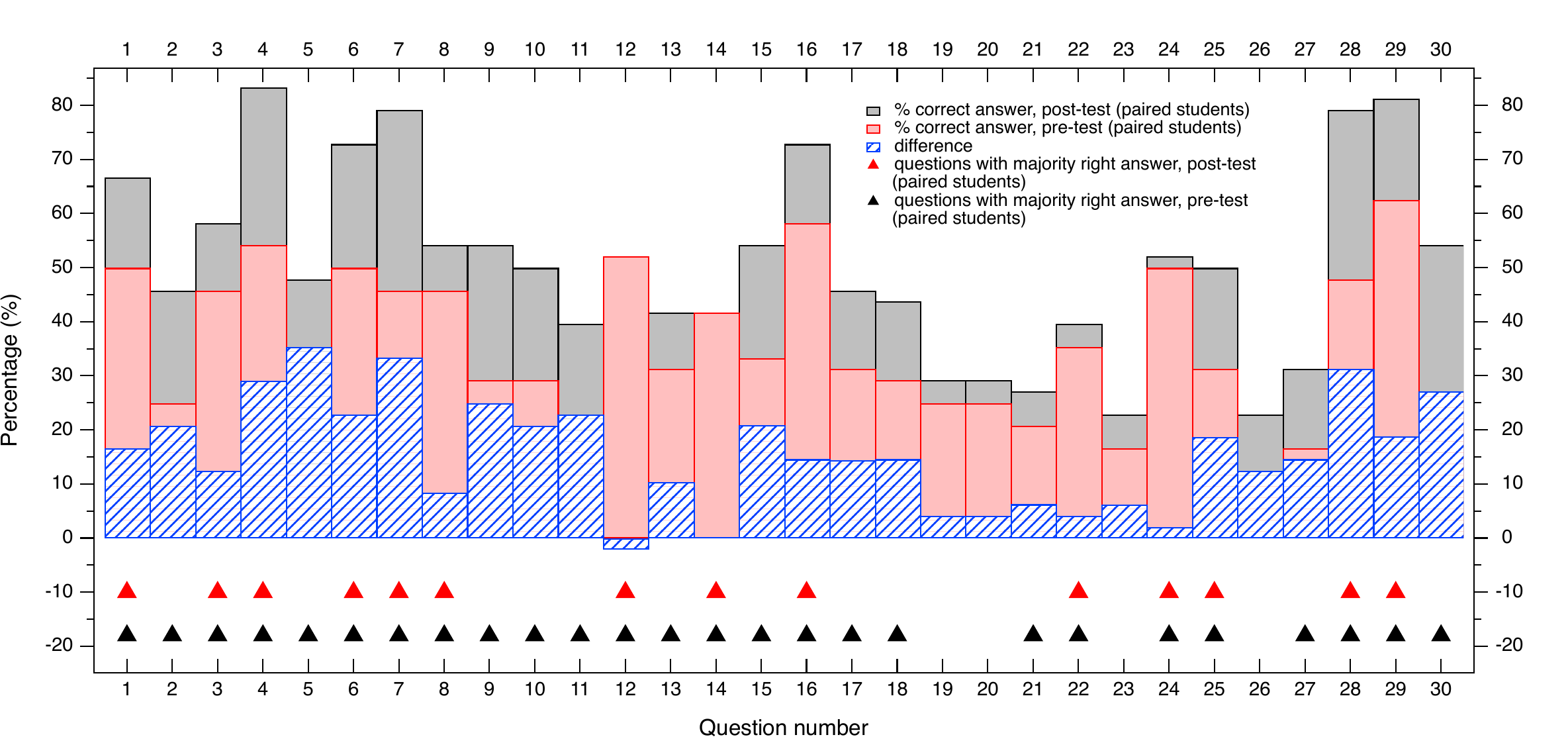}}
    \caption{Top: Correct answer analysis for all the students ($N=256$) who wrote pre- and post-tests. Bottom: Correct answer analysis for all the paired students ($N=48$) who wrote pre- and post-tests.}
    \label{fig:post-pre-breakdown}
\end{figure}

\section{Results \& Analyses}
\label{sec:3}
\subsection{Question Break Down, Means and Gains}
\label{sec:3means}

\par In this section we first present some analyses of the distribution of the scores for students in 
the pre- and post-tests ($N=256$), starting in the left panel of Fig. \ref{fig:HistMarks} (where the 
right panel is for the paired data, $N=48$). The difference in distributions (pre- compared to 
post-test) have well defined shifts indicating a gain, particularly for the paired data. In Table 
\ref{table-marks-stats} the results for the pre- and post-test scores can be seen. The gain of $G=0.24$ 
was calculated from the paired data and is the expected gain for a standard physics course 
\cite{Hake:1998}. The difference in mean scores was checked via the paired samples {\it t}-test 
($N=48$) and was {\it not} due to random fluctuations at the $95\%$ confidence level ({\it p}-value $= 
0.000002614$, two-tailed), implying we found a statistically significant difference in the means (pre- 
and post-test).\footnote{The means for the pre- or post-tests groups used an independent samples {\it 
t}-test and was found to be not due to random fluctuations at the $95\%$ confidence level; {\it 
p}-value $=0.00001182$, two-tailed.} 

\par This relationship between pre- and post-test scores can also be seen in Fig. 
\ref{fig:correlation-marks}, where the correlation was found to be moderate and positive. It should 
be noted that the gain from the pre-test mean, as compared to the post-test mean, is not used to 
determine the gains \cite{Hake:1998}, although we have performed a question by question breakdown. In 
terms of numbers, Pearson's correlation coefficient, in Fig. \ref{fig:correlation-marks}, gives a 
moderate positive correlation of $r=0.504$ with {\it p}-value $= 0.000261<\alpha$ at the $95\%$ 
significance level ($\alpha<0.05$).

\par As student ideas may not be clearly understood, 
we can identify them through analyses of question by question responses. Fig. 
\ref{fig:post-pre-breakdown} indicates the percentages of students who correctly answered each question
in the pre- and post-tests, with the differences also shown (diagonal hatching/blue). A similar 
schematic is shown for the paired data (see bottom panel Fig. \ref{fig:post-pre-breakdown}) for 
comparison. We note here the presence of negative gains for certain questions, which are indicative of 
poorer performance in the post-test. This ``loss'' is especially pronounced in Fig. 
\ref{fig:post-pre-breakdown}, top panel, for questions 14, 20, 22, and 24; this also arises in the 
paired data of Fig. \ref{fig:post-pre-breakdown} for question 12, bottom panel; the negligible gains in
question 21 for the former and questions 14 and 24 for the latter are also worth mentioning. The 
concepts assessed by these questions are standard topics such as projectile motion (questions 12 and 
14), as well as kinematics and Newton's second law (questions 20-24). However, these questions possibly
exploit scenarios with which students are unlikely to have had personal experience with (i.e. objects 
fired from cannons and motion in deep space) and visual tools like ticker-tapes and displacement-time 
graphs. In doing so, they ensure that students respond based on intuition gained from their mechanics 
course rather than empirical evidence gathered from daily life.

\par Finally, poorer post-test performance in these more conceptual questions may demonstrate that 
students are not confident in their ability to apply their knowledge to unfamiliar scenarios. This may 
be a consequence of superficial learning or dependence on preconceived ideas rather than physics. The 
presence or development of ``misconceptions" may also have come into play. Additionally, we note that 
these questions might indicate an issue with language ability, a concern shared by other studies 
conducted in regions where English is not the first language of most students 
\cite{Alinea:2015,Bani_Salameh:2016a}. 

\subsection{Polarising Questions}
\label{sec:3pol}
\par
\begin{figure}[ht]
    \centering
    \scalebox{0.25}{
    \includegraphics{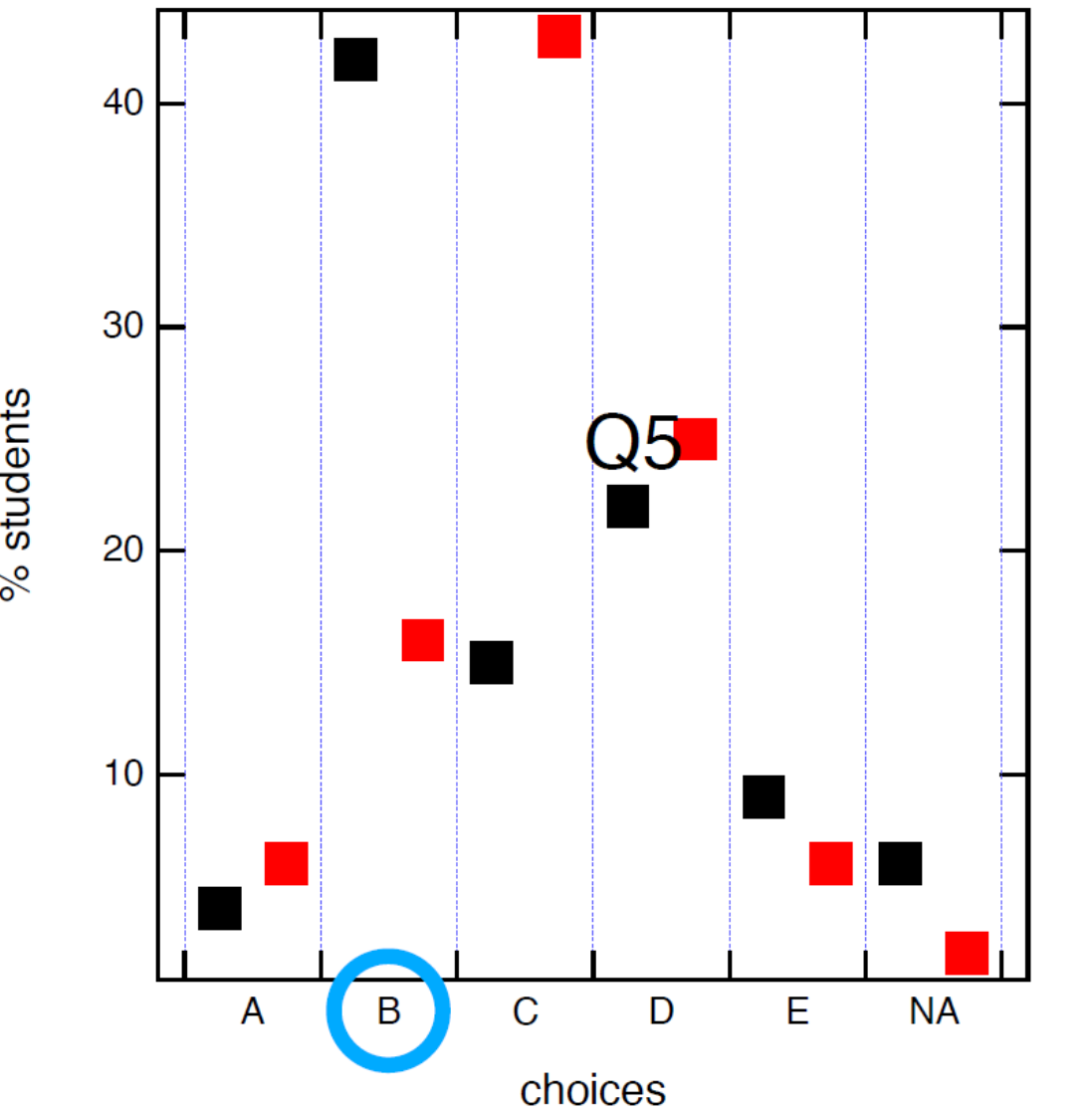}}
    \scalebox{0.25}{
    \includegraphics{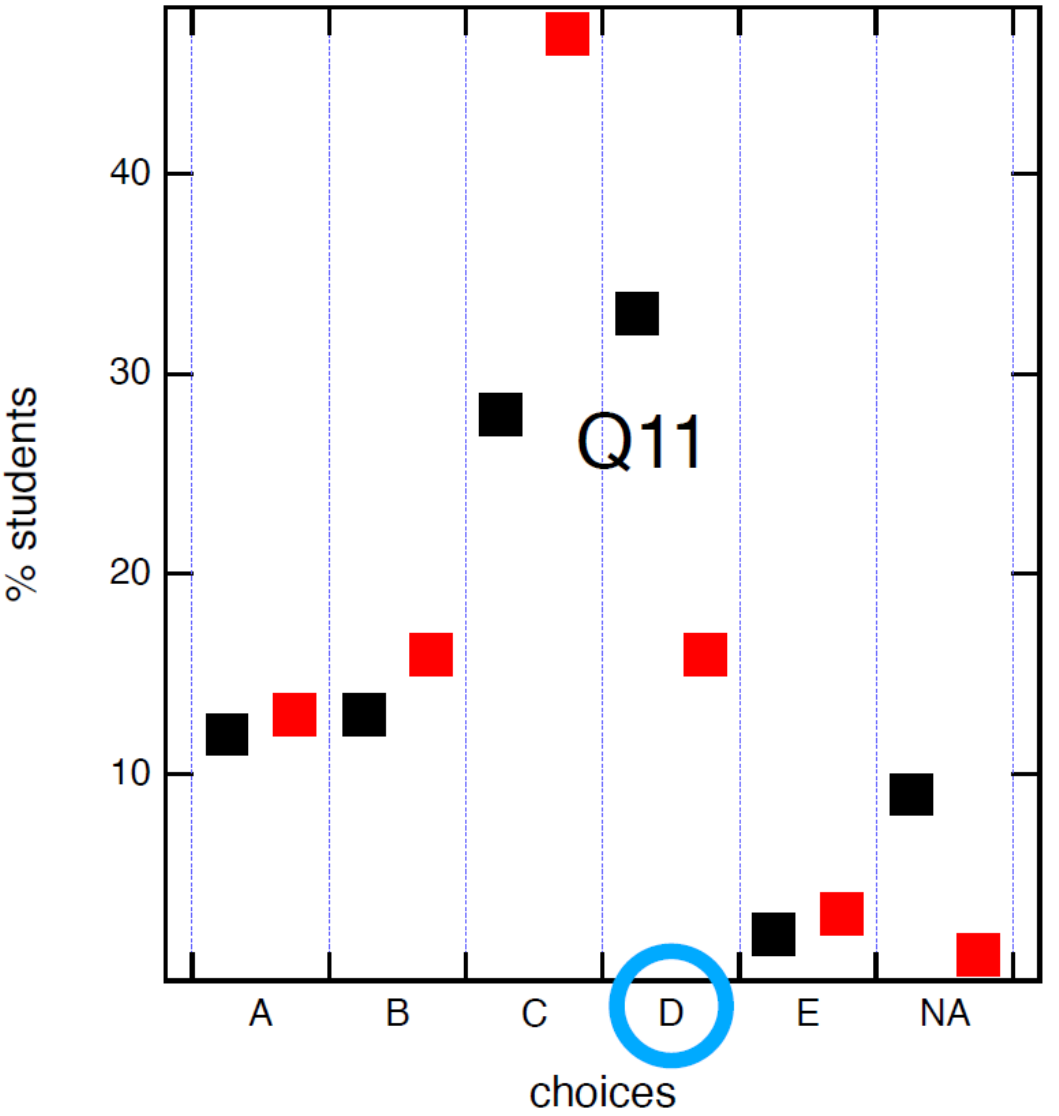}}
    \scalebox{0.25}{
    \includegraphics{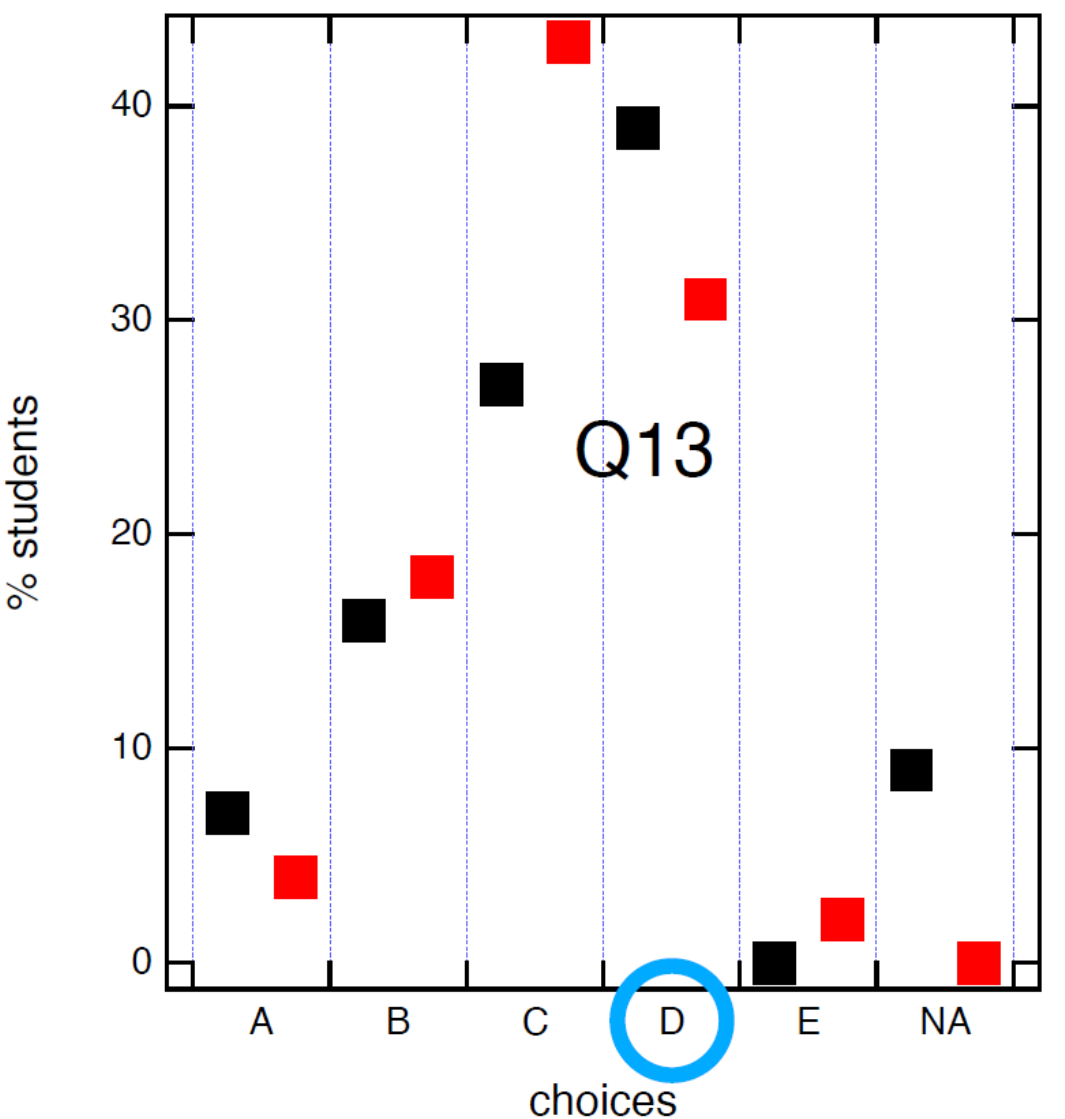}}
    \scalebox{0.25}{
    \includegraphics{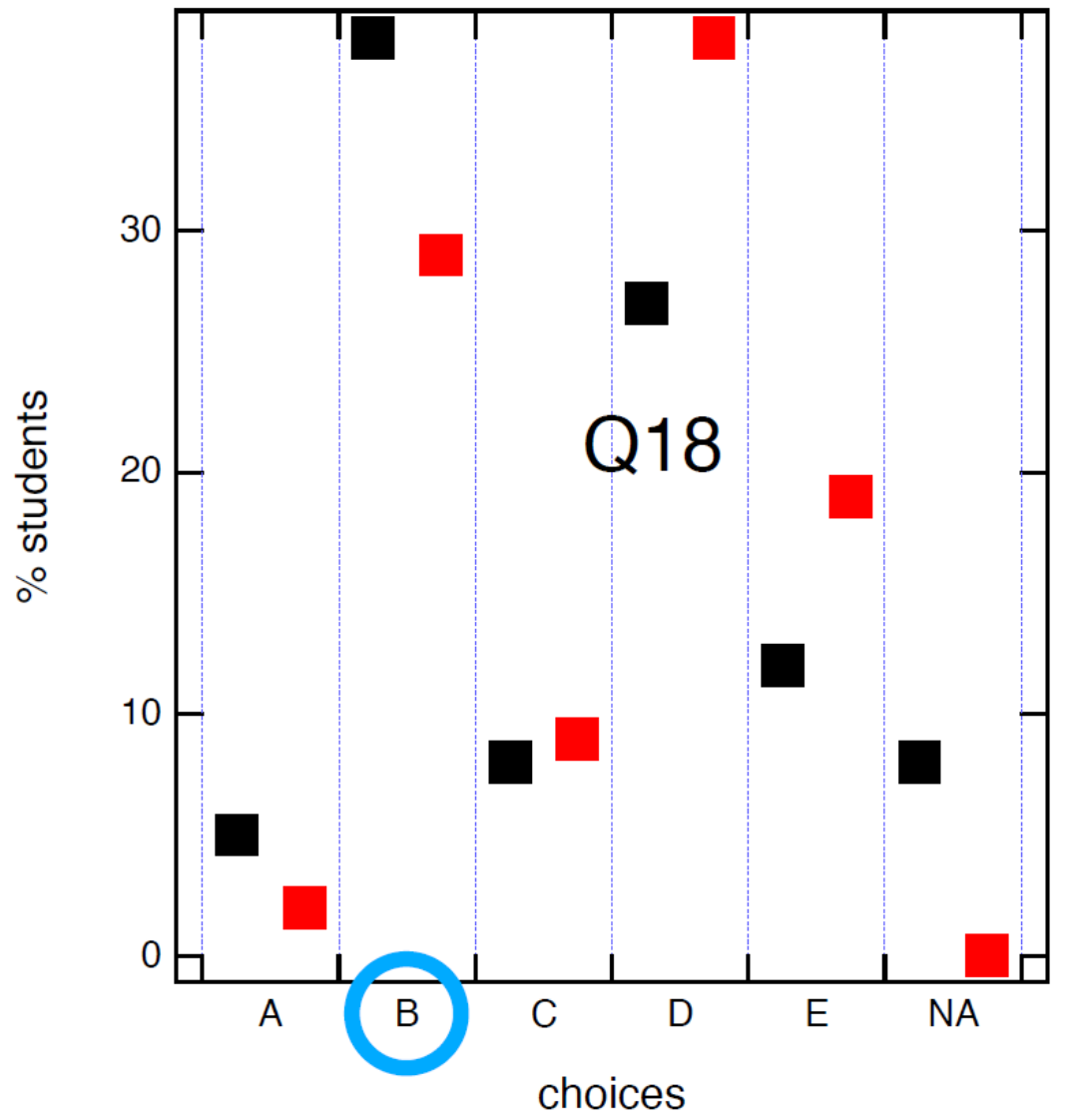}}
    \scalebox{0.25}{
    \includegraphics{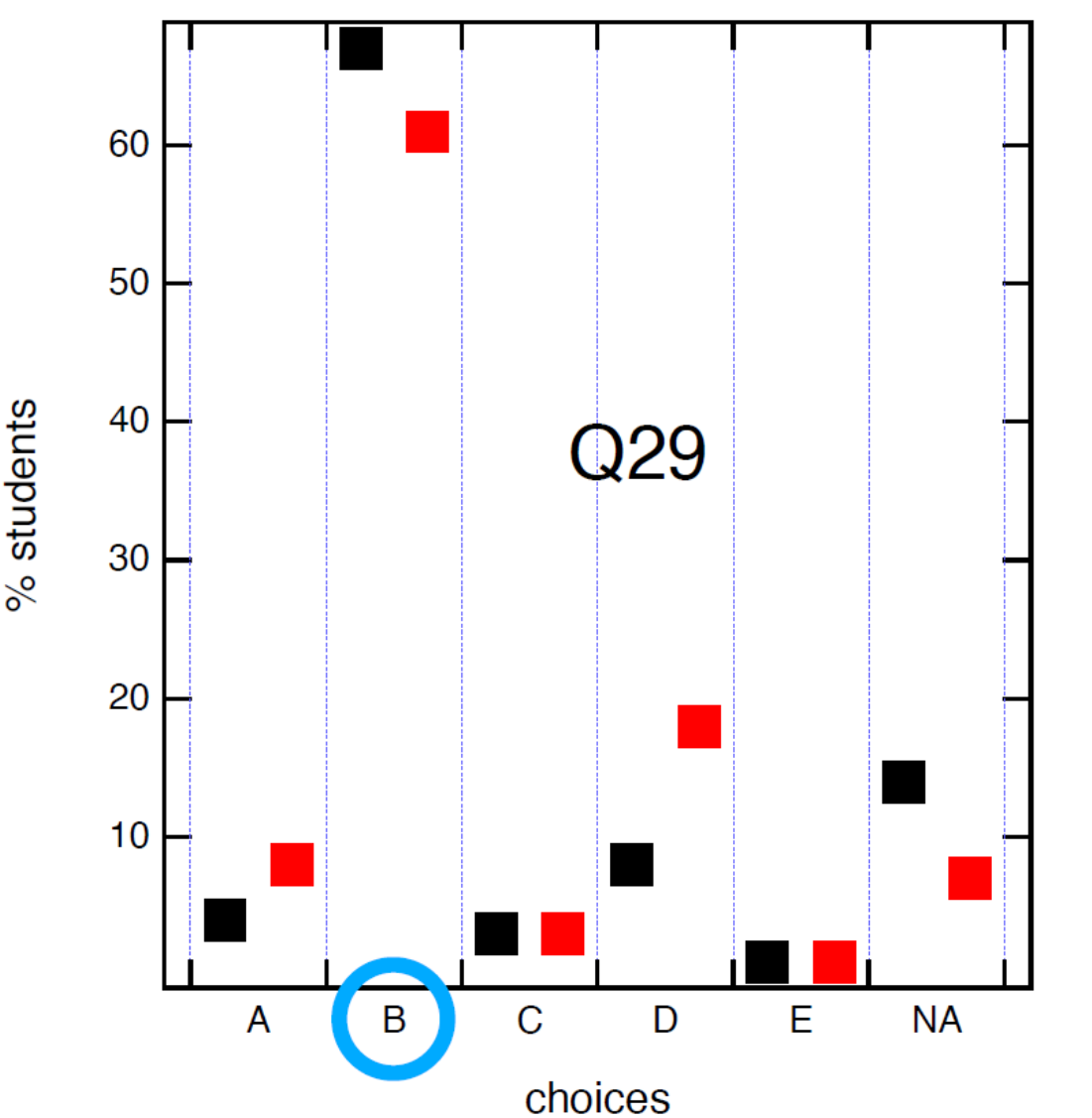}}
    \scalebox{0.25}{
    \includegraphics{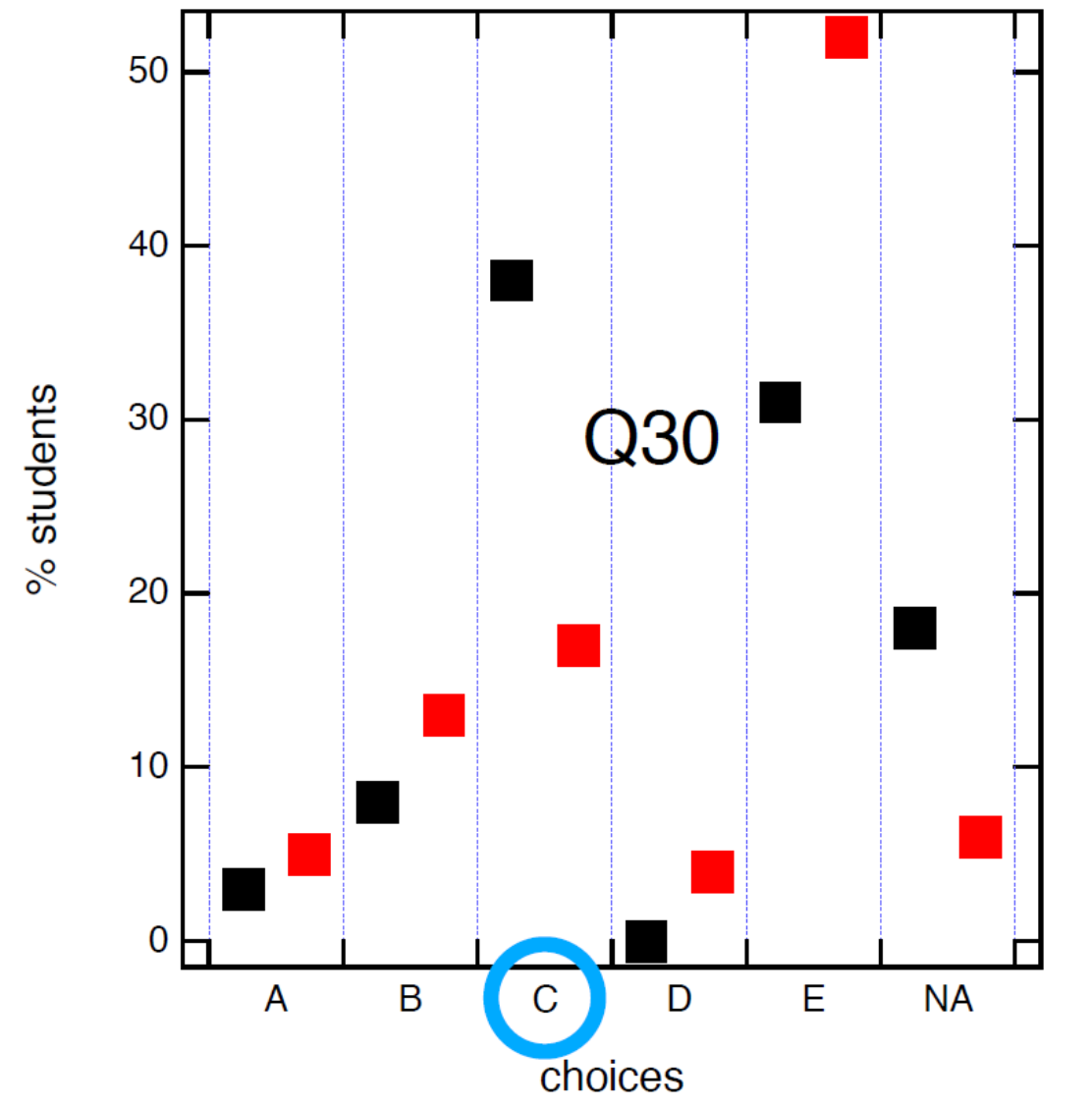}}
    \caption{Distribution of student answers for Questions  $5, 11, 13, 18, 29$ and $30$, pre-(red) and post-(blue), for all $N=256$ students. NA stands for ``no answer'', while the five options are labelled from A to E. Correct response circled.}
    \label{fig:polarisation_All}
\end{figure}

\begin{figure}[ht]
    \centering
    \scalebox{0.24}{
    \includegraphics{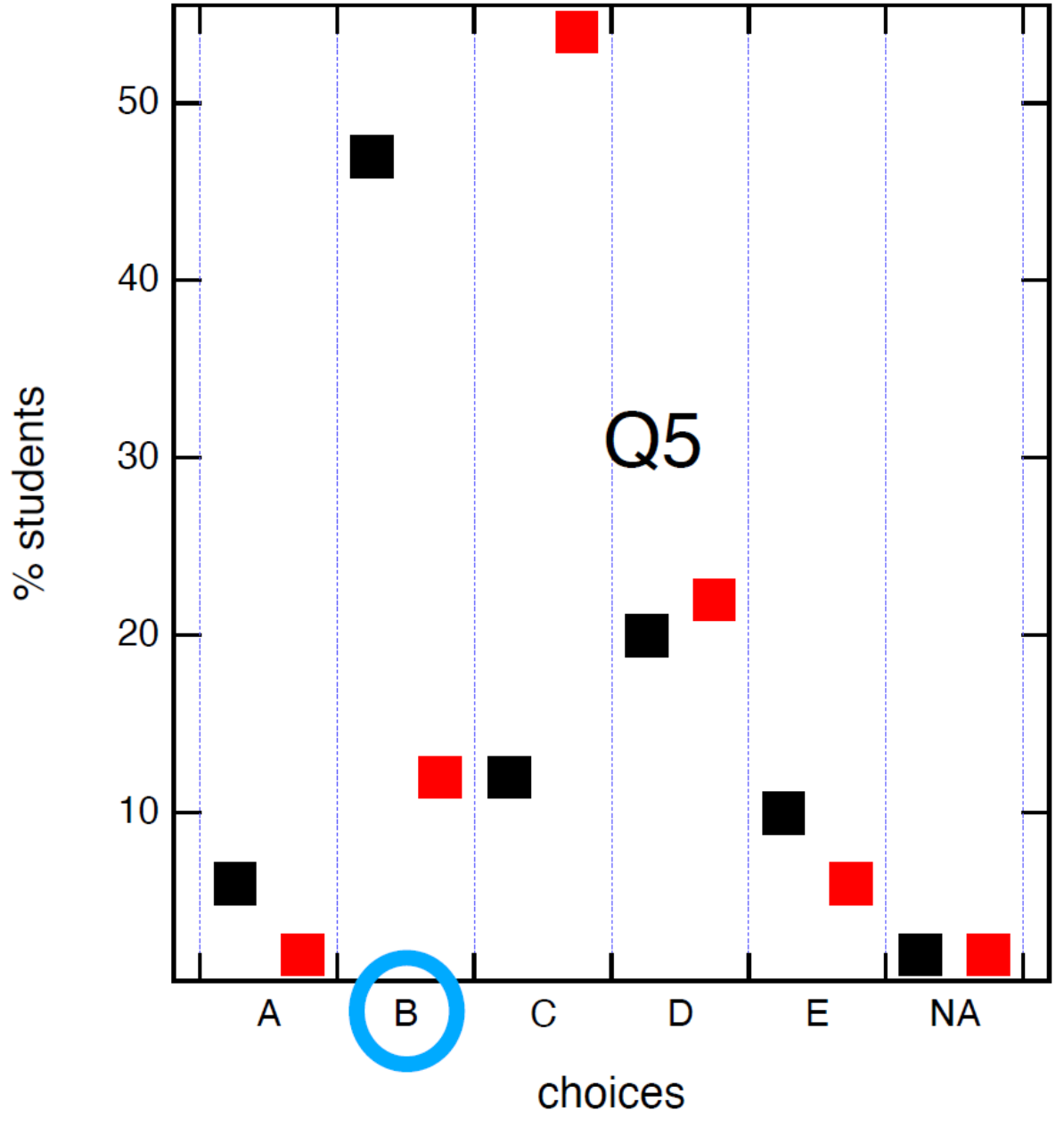}}
    \scalebox{0.24}{
    \includegraphics{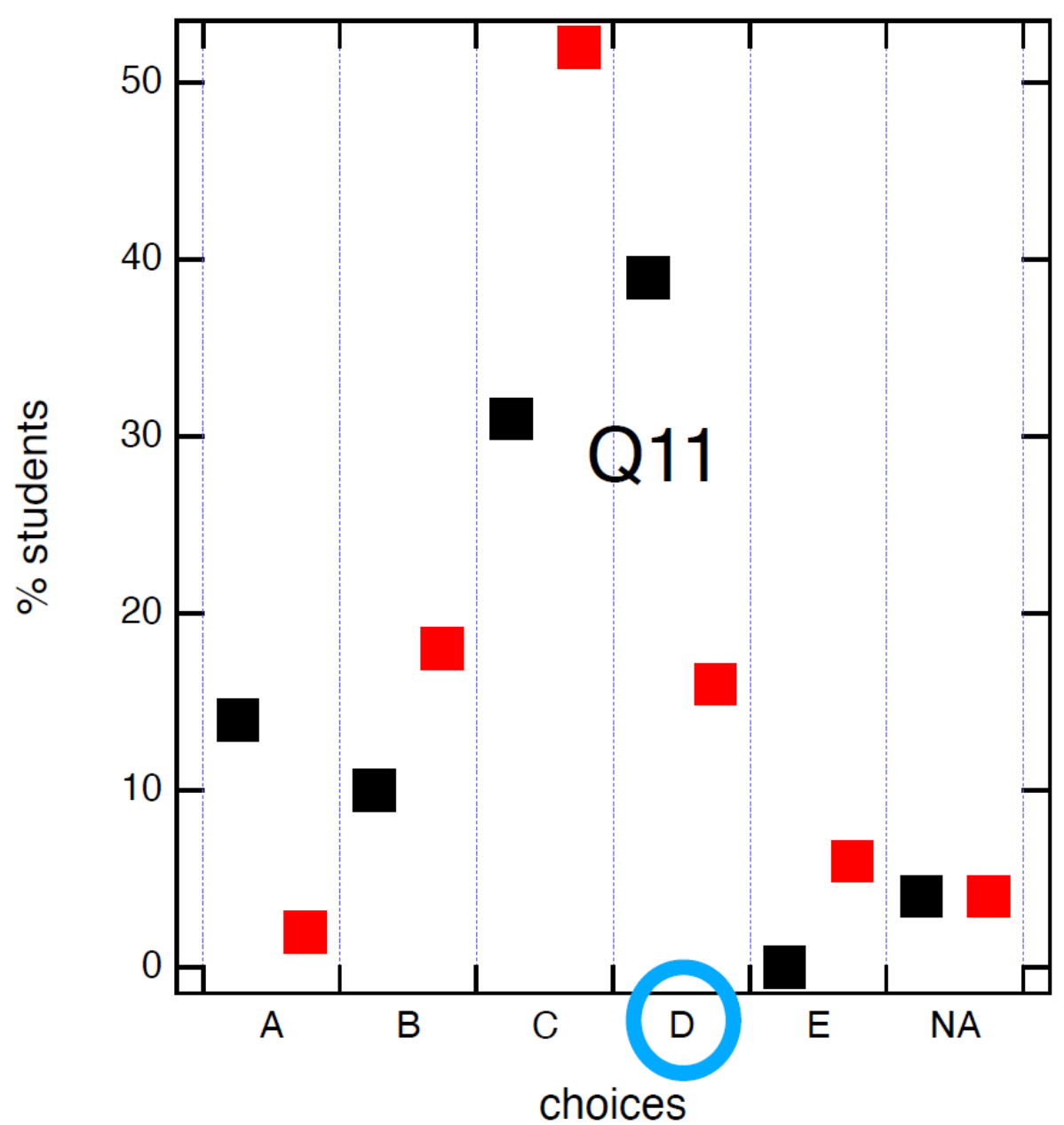}}
    \scalebox{0.24}{
    \includegraphics{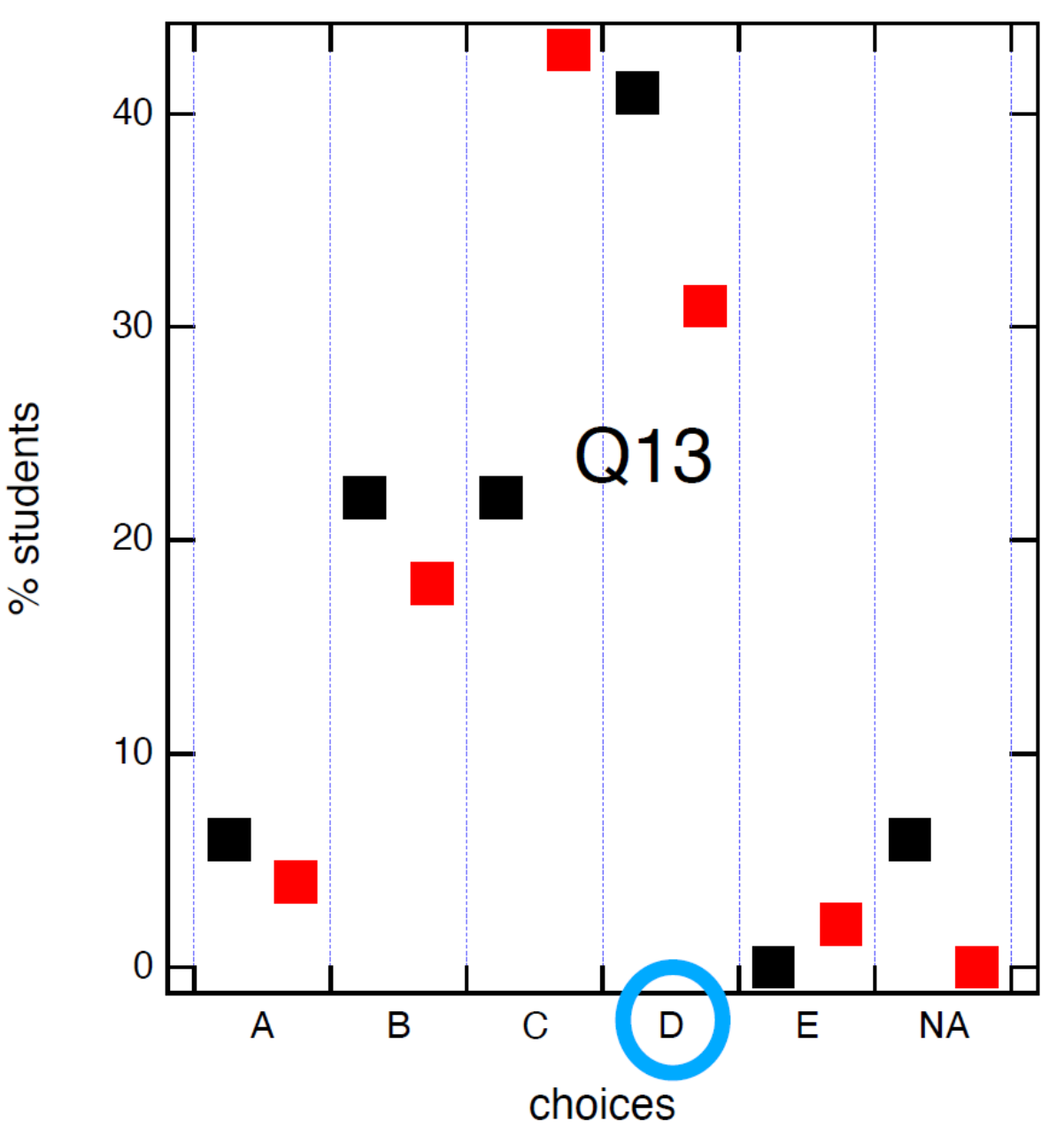}}
    \scalebox{0.24}{
    \includegraphics{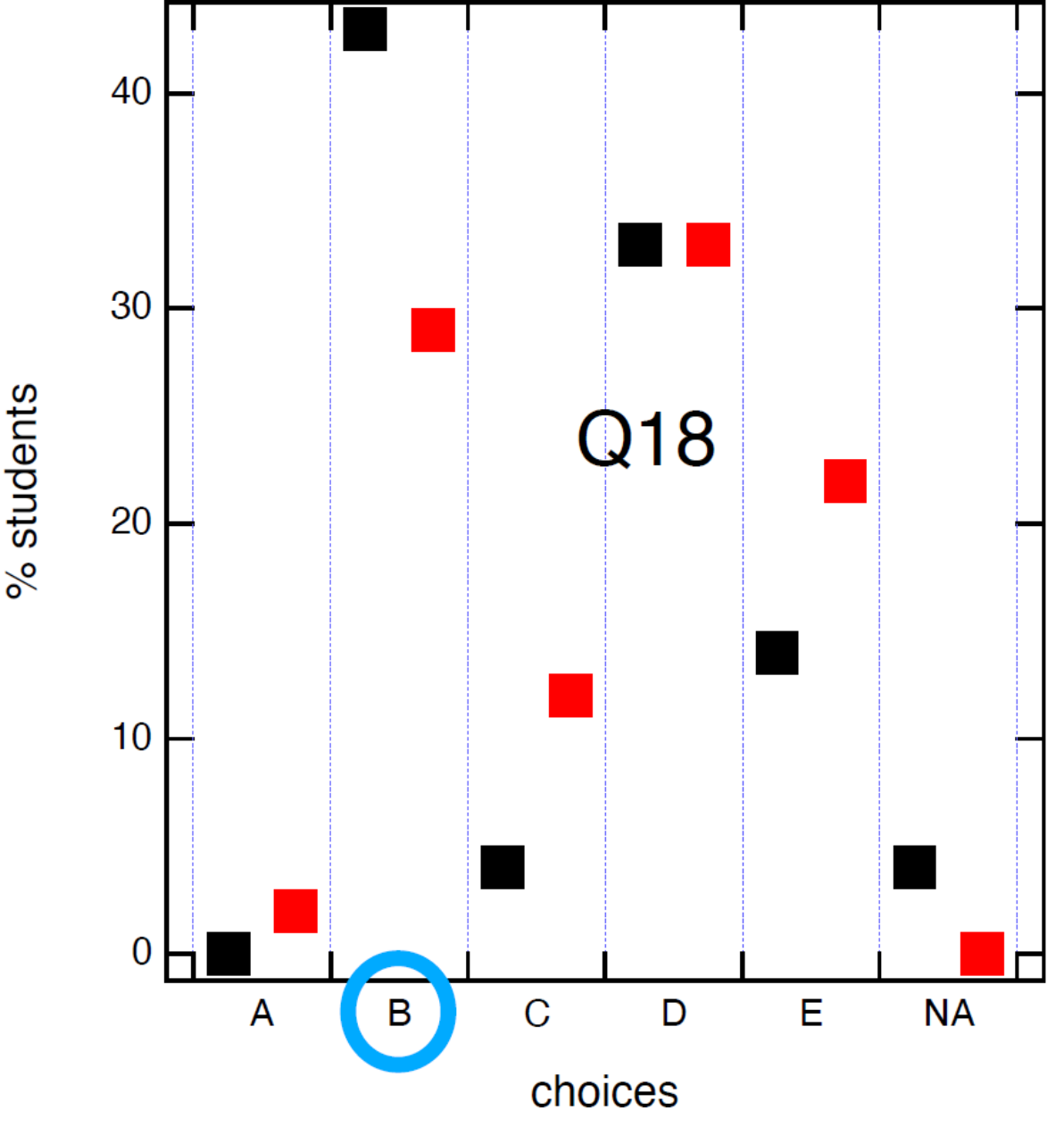}}
    \scalebox{0.24}{
    \includegraphics{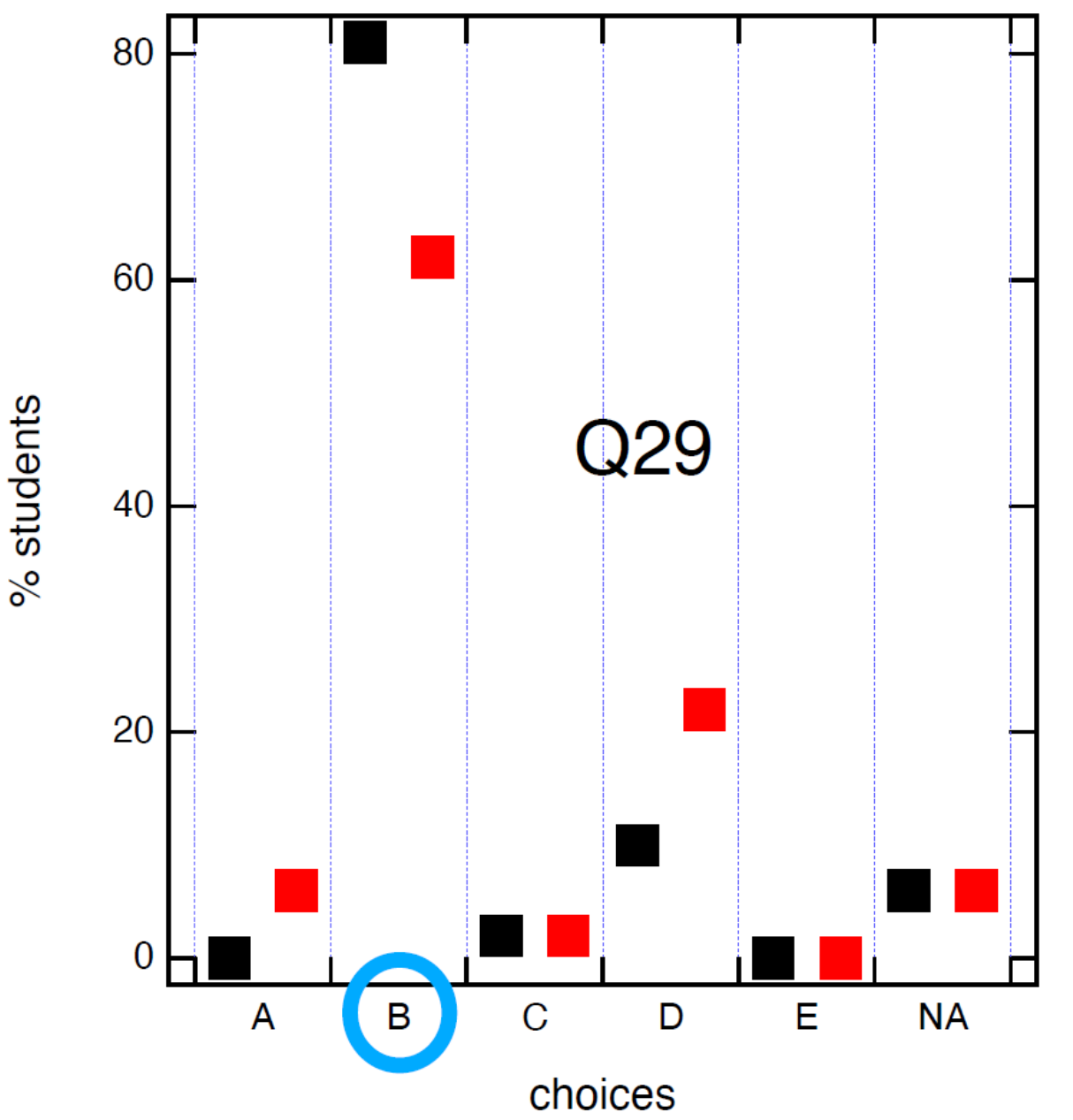}}
    \scalebox{0.24}{
    \includegraphics{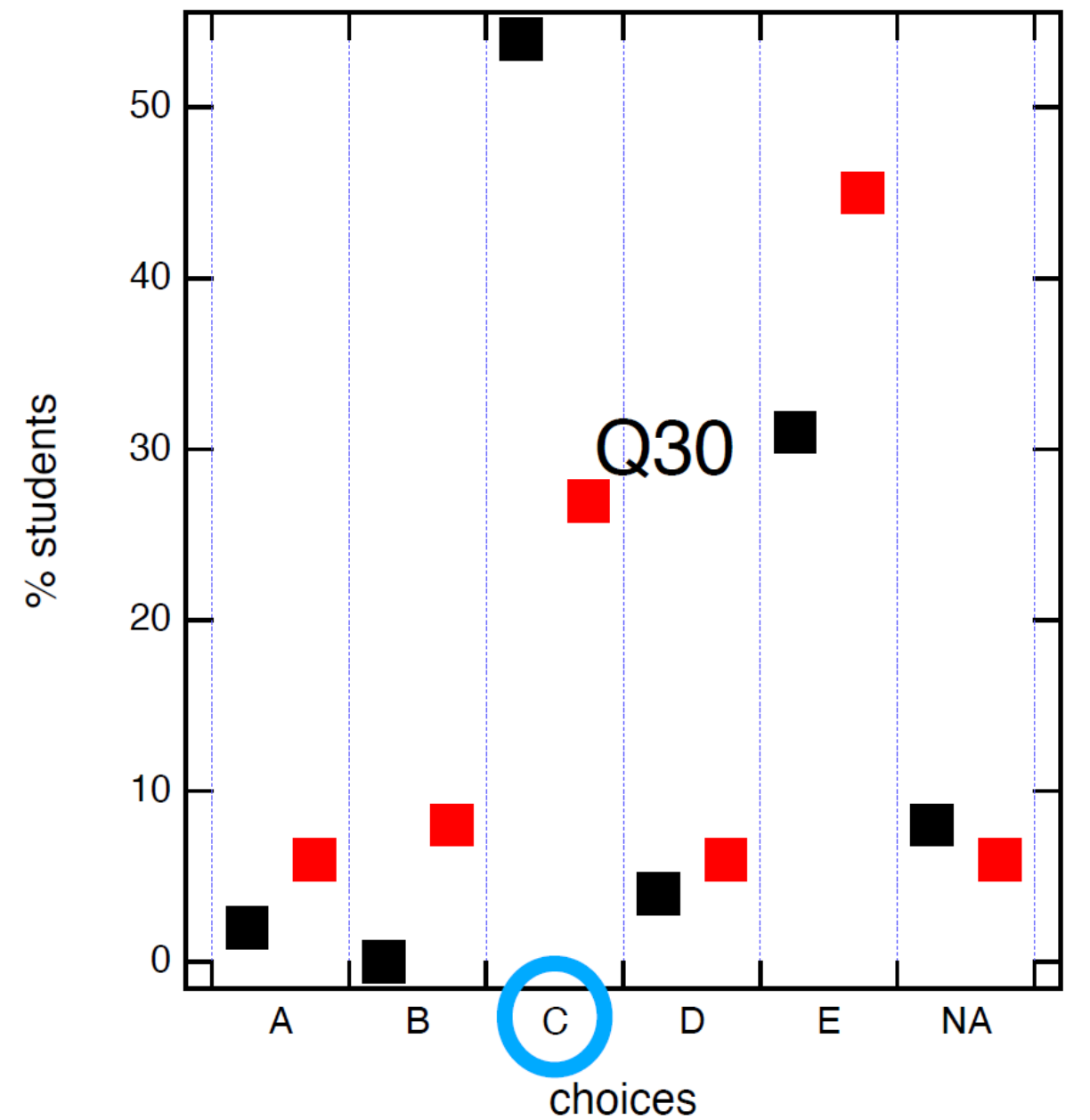}}
    \caption{Paired distribution of student answers for questions  $5, 11, 13, 18, 29$ and $30$, pre-(red) and post-(blue), tests for the paired group ($N=48$). The same conventions are used as in Fig. \ref{fig:polarisation_All}.}
    \label{fig:polarisation_PAIRED}
\end{figure}

 As discussed in the introduction we can also analyse particular sets of questions which can conceptually lead to polarising choices \cite{Alinea:2015}. In Figs. \ref{fig:polarisation_All} and \ref{fig:polarisation_PAIRED} we can see the effect of the polarising questions: $5, 11, 13, 18, 29$ and $30$ in the FCI, e.g., see Ref. \cite{Alinea:2015}. A similar pattern emerges for the cohort at UJ, where there clearly appears to be a subset where asides from the correct answer there is another polarising choice, and apart from Q18 in the ``paired'' data, Fig. \ref{fig:polarisation_PAIRED}, we find the same dominant incorrect (polarised) response \cite{Alinea:2015}.

\par It may be that certain ``misconceptions" drive this polarisation. For example, consider question 5, where the dominant answer of C can be read from the pre-test data of Figs. \ref{fig:polarisation_All} and \ref{fig:polarisation_PAIRED}. This answer claims that the motion of a ball is driven by gravity as well as ``a force in the direction of motion'', which indicates the common misconception that motion requires an active force. The presence of a force in the direction of motion is favoured also in answers $11$C, $13$C, and $18$D, as well as implied in $30$E. Though there is a general decrease from pre- to post-test in the selection of these erroneous answers, the post-test data of Figs. \ref{fig:polarisation_All} and \ref{fig:polarisation_PAIRED} suggest that these ``misconceptions" can be difficult to alleviate, as we have found at UJ.

\par Such observations connect well to the work of Bani-Salameh \cite{Bani_Salameh:2016b} and others; we shall interrogate these ideas in greater depth in a future work with data from 2020-2021, see Sec. \ref{sec:4}. It can certainly be inferred that there are subsets of incorrect answers where ````misconceptions"" in students' understanding consistently leads to the same kind of wrong answer \cite{Alinea:2015}. 

\begin{table}[ht]
\centering
    \begin{tabular}{|l|ccc|}
    \hline
    \multirow{2}{*}{Group} & Female  & Male  & Difference \\
       & $n=16$ & $n=32$ &  \\
    \hline
    \hline
       & Mean (SD)    & Mean (SD)  & (Male-Female) \\
    Pre-test(\%) & 31.0 (12.3) & 37.0 (19.4) &  6.0 \\
    Post-test(\%) & 56.9 (25.2) & 47.6 (20.5) & -9.3 \\
    Gain(\%)   & 38.0  & 17.3 &   -20.7  \\
    \hline
    \end{tabular}
    \caption{Paired FCI results (in percent) for female and male participants ($N=48$).}
    \label{table_gen}
\end{table}
\begin{figure}[ht]
    \centering
    \scalebox{0.5}{
    \includegraphics{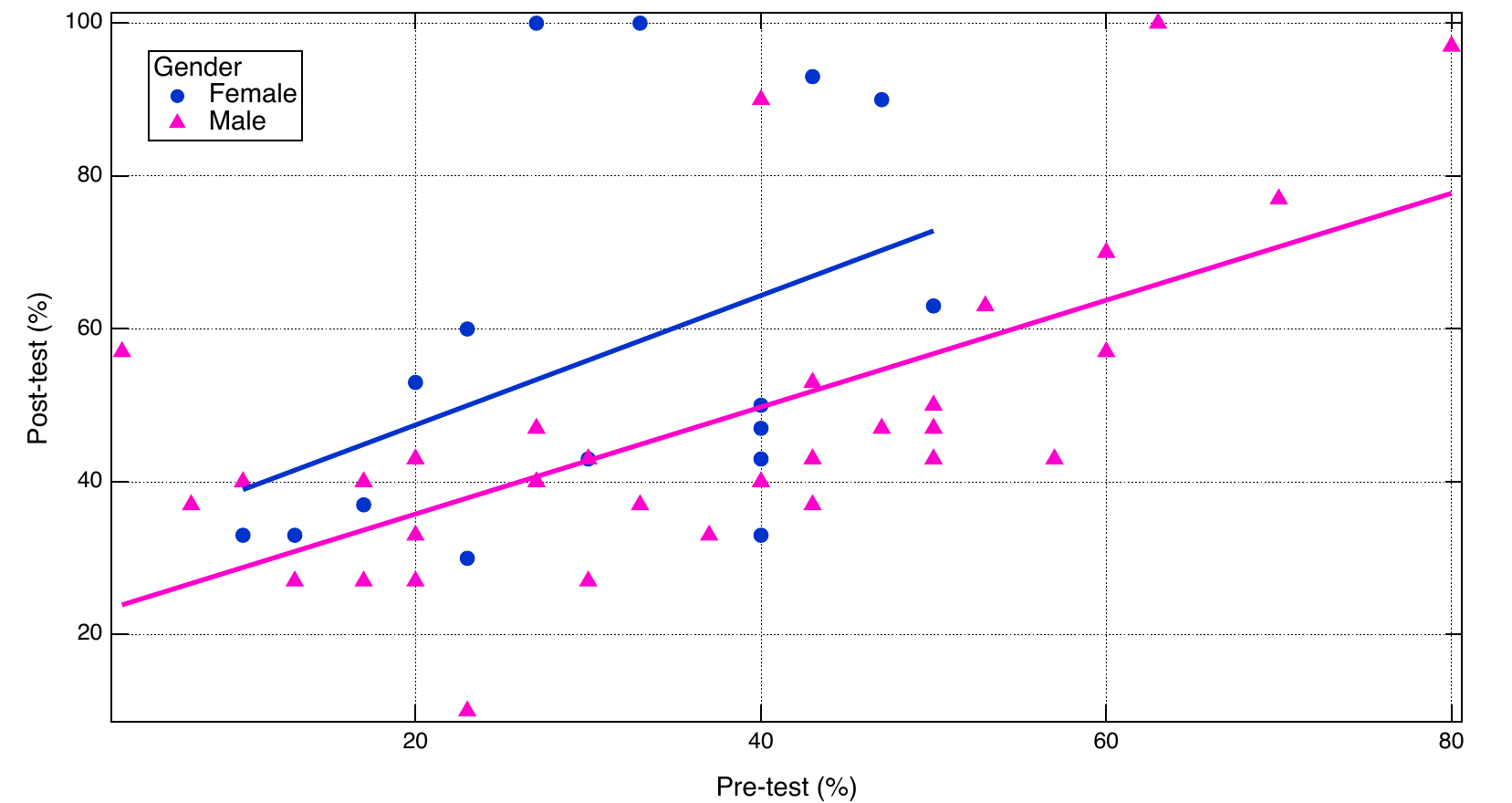}}
    \caption{Correlations for combined scores in terms of gender.}
    \label{fig:correlation-gender}
\end{figure}

\subsection{Possible Gender Gains}
\label{sec:3gend}

\par In this section we now analyse the differences between the male and female participants who sat 
for both pre- and post-tests (paired). In Table \ref{table_gen} the means for female and male 
participants are presented and it clearly appears that female participants have performed better in the
FCI as compared to male participants. Interestingly, and it has also been found by Alinea \& Naylor 
\cite{Alinea:2017}, the table shows that although male participants did better in the pre-test, female 
participants had a higher average in the post-test. 

\par To try to verify these results further, where due to the average number of participants in each 
group being $24$, we have (in Appendix \ref{app:A}) performed multiple statistical tests to confirm if 
this difference in means is statistically significant. From this we have found that at the $95\%$ 
significance level we can neglect the null hypothesis. We emphasize that besides parametric tests for 
normal distributions we have also performed a non-parametric Wilcoxon test that led to a statistical 
difference in the medians, see App. \ref{app:A}.

\par Besides the means, in comparison to Fig. \ref{fig:correlation-marks}, in Fig. 
\ref{fig:correlation-gender} the correlation for female and male groups was found to be mild and 
positive with: $r_F=0.417$; {\it p}-value $= 0.05418$ and $r_M= 0.662$; {\it p}-value$=0.00001816$, 
respectively. Clearly, there is a more reliable correlation for the male cohort with {\it p}-value $< 
0.05$. Finally, this can be compared to the combined correlation (independent of gender, $N=48$) where 
$r = 0.504$; {\it p}-value$=0.02757$ in Fig. \ref{fig:correlation-marks}.

\par The reason for the {\it apparently} higher gain for the female part of the cohort might be due to, 
as found by Sadler and Tai \cite{Sadler:2001} (also see Adams {\it et al} \cite{Adams:2006}), 
professor-to-gender matching to student gender was second only to the quality of high-school physics 
course in predicting students' performance in college physics. It may be worth mentioning that at UJ, 
during the 2020 academic year, a female instructor was the senior academic for the mechanics course. As
we discuss in Sec. \ref{sec:4}, we will leave these preliminary results for a follow-up work, with more
years of data.

\section{Concluding Remarks}
\label{sec:4}

\par In this article we have used the Force Concept Inventory (FCI) to look at the conceptual 
understanding of  a large cohort of physics/engineering students at the University of Johannesburg (UJ) 
during the 2020 academic year, see Sec. \ref{sec:3means}. Mid-semester, UJ went into lock-down and 
students then switched from a 
traditional lecture format to online platforms, yet this led to the very informative scenario where we 
have found no overall drop 
in conceptual gains ($G=0.24$). This is reminiscent of what happened with regards to the Christchurch 
Earthquakes (2010-2011) which led to the closure of various high schools, where although a minority had 
negative impacts there were were many positives \cite{Beagle:2017}.\footnote{In terms of physics 
performance during COVID see Refs.  
\cite{White:2020,klein:2020,fox:2020}.} This was further established through the comparison of 2019 and 
2020 semester marks at UJ (see Table \ref{table-marks-2019-2020-comp}) where we found no appreciable 
drop in marks. {\color{black}{The extent to which the teaching style used in UJ's Physics Department, as it involves active learning such as in-class discussions and problem-solving, weekly tutorials, etc., potentially contributed to this consistency. Note that this is worth probing through deeper comparisons with other investigations into forced transitions to online learning.}}

\par In Sec. \ref{sec:3pol} we looked at a subset of questions where a polarisation of 
choices occurred, in 
that either the correct answer or one main incorrect answer dominated the post-test responses 
\cite{Alinea:2015}. We found similar patterns for the students at UJ, which followed a very similar 
pattern to the data found in Refs. \cite{Alinea:2015,Alinea:2017,Alinea:2020}. The importance of these 
questions relates to the fact that they ask the student to be able to understand certain particular 
concepts in physics, such as circular motion and motion requiring a force.

\par In Sec. \ref{sec:3gend} we looked at a possible out-performance of female students on the overall 
gain in
the FCI. As was also found by Alinea \& Naylor \cite{Alinea:2017}, although the male group started with 
a higher average pre-test score, their gain was less. As mentioned earlier, the main course lecturer was
female, which may lead to professor gender matching in this cohort \cite{Sadler:2001}. Although we 
rigorously checked 
that the difference in means was statistically significant (at the $95\%$ confidence level, see App. 
\ref{app:A}) we will report on a larger cohort inclusive of $2021$ in forthcoming work, where we hope to
investigate if other factors may be related to the diligence of female students and socioeconomic 
factors.

\par The article has raised some questions, such as why the general performance for the group of paired 
students was higher than those who took either of the pre- or post-tests. This is often  due to the fact that 
students who are diligent are more likely to take both pre- and post-tests. Usually, overall gains are taken 
only from paired data, which is then used to compare to other cohorts and institutions. However, the question 
of using ``unpaired'' pre- and/or post-test data sets in some form has not really been investigated in the 
literature (see however, Ref. \cite{Bani_Salameh:2016a}) and we also intend to comment more on this issue in 
future work. 

\par As for other possible directions to investigate, besides extending the FCI to further years,
which also have been disrupted by more COVID lockdowns, we intend to look at matriculation 
results in order to establish 
correlations between the FCI and high school exit grades in physics, maths and English scores. 
In the case of UJ, the first language of the students enrolled in the UJ FEBE (average 
2015-2019): English 14.3$\%$; Isixhosa 5.9$\%$; African 1.5$\%$; Other 78.3$\%$ 
\cite{FEBE2019AnnualReport}. 
This relates to comments by Bani-Salameh \cite{Bani_Salameh:2016a}, and also work performed by 
Alinea \& Naylor \cite{Alinea:2015}, in relation to performance on the English version of the 
FCI, where English is not 
the student's first language necessarily. It might also be interesting to look at correlations 
between the FCI and cognitive reflective tests \cite{Alinea:2020}.

\par {\color{black}{Finally, it is important to recognise the limitations of the FCI, where as a multiple-choice tool  designed to probe ``misconceptions" rather than context-based understanding \cite{diSessa2018_KIP}, much of the insights obtained through its use depend on the researcher's interpretation of students' answers. It would be interesting to deploy the FCI with the added expectation that students justify each of their answers. This would allow for a more nuanced and less biased analysis that goes beyond understanding ``misconceptions". An additional means of expanding our analysis of first year physics pedagogy would be to integrate principles such as ``knowledge-in-pieces'', which rests upon the notion that ``knowledge depends on context'' \cite{diSessa2018_KIP}.}}

\par {\color{black}{On a related note and to unpack fully the issue of whether or not the negative (or zero) gains could be due to  limitations in the FCI, it would be worthwhile to get students to give written explanations for their choices. This would again advocate the use of technology in using conceptual testing. Hence, given the sometimes necessary switch to online platforms, such as Blackboard \cite{blackboard2016} etc., as well as obtaining question explanations, we would also able to study the engagement of students within this online learning environment (through their attendance and marks on continuous assessments).} This would also include the time taken to complete various assessment tasks.}


\section*{Acknowledgements}
 EC and ASC are supported in part by the National Research Foundation of South Africa (NRF). AC is grateful 
 for the support of the National Institute for Theoretical Physics (NITheP), South Africa. This study was 
 done in compliance with the South African Protection Of Personal Information (POPI) Act, where all student 
 data (including personal data) was anonymised and collected as part of the University of Johannesburg's 
 physics course’s online assessment platform.
 We would like to thank all staff and students who took part in this study. WN would like to thank useful 
 discussions with Margaret Marshman, University of the Sunshine Coast. The authors are also grateful 
 to Allan L. Alinea (University of the Philippines) for his useful comments. 


\begin{appendices}

\section{Statistical Analyses of Differences in Gender Means}
\label{app:A}

\par In this appendix we look at the differences in gender means for paired data ($N=48$, comprising of $16$ female participants and $32$ male participants). We found that 
the mean of the gains for female participants ($\mu_G^F=0.38$) was greater than the mean for male participants ($\mu_G^M=0.17$). However, to clarify if the difference is 
purely a random fluctuation, we performed the following tests, using R \cite{Rprogram}, at the $95\%$ significance level ($\alpha<0.05$):

\begin{itemize}
\item[(a)] The {\it t}-test for independent samples (and unequal variances) had a {\it p}-value: $0.02757<\alpha$ for a single tail (directional difference $\mu_G^{F} - \mu_G^{M}>0$). 

\item[(b)] A two-way ANOVA with replication led to an $F$-statistic: $4.5107$ and {\it p}-value: $0.03909<\alpha$. 

\item[(c)] A linear regression analysis also led to an $F$-statistic: $4.511$ and {\it p}-value: $0.03909<\alpha$.

\item[(d)] A non-parametric two-sample Wilcoxon test led to medians of $0.2496296$ and $0.1602564$ for female participants and male participants, respectively, with a 
$W$-statistic of $W = 341$ and {\it p}-value $= 0.03223<\alpha$.  
\end{itemize}

\par It may be worth mentioning that two-way ANOVA with replication was unbalanced, as the two group sizes were different ($16$ and $32$, respectively). However, we were able
to double-check the results obtained by converting gender to a dichotomous variable (Female$=1$, Male$=0$) and used a linear regression. At the $95\%$ significance level (or $\alpha<0.05$) we can reject the null-hypothesis whenever the {\it p}-value $< 0.05$.

\par In conjunction with the {\it t}-test, a two-way ANOVA and a linear regression analysis both agree at the $95\%$ significance level, and suggest that there is a 
statistically significant difference between the means of female participants and   male participants. This was further confirmed in item (d), where we performed a 
non-parametric test (using medians) and found the critical value to be $p = 0.03223<\alpha$. These findings indicate a {\it real}  difference in gender (for this group) with 
female participants having better gains than male participants, even though male participants started with a higher average pre-test score.

\end{appendices}

\section*{References}
\bibliographystyle{ieeetr}
\bibliography{Education}



\end{document}